\documentclass[5p]{elsarticle}

\usepackage{epsfig}
\usepackage{amsmath}
\usepackage{amssymb}
\usepackage{color}
\usepackage{array}

\newcommand{\slabel}[1]{\label{sec:#1}}
\newcommand{\elabel}[1]{\label{eqn:#1}}
\newcommand{\flabel}[1]{\label{fig:#1}}

\newcommand{\cref}[1]{\ref{cpt:#1}}
\newcommand{\sref}[1]{\ref{sec:#1}}
\newcommand{\eref}[1]{\ref{eqn:#1}}
\newcommand{\fref}[1]{\ref{fig:#1}}

\newcommand{\listBegin}{\begin{tabular}{cp{4.5in}}}
\newcommand{\listEnd}{\end{tabular}}

\newcommand{\matrixBegin}[1]{\left[\!\!\left[ \begin{array}{#1}}
\newcommand{\matrixEnd}{\end{array} \right]\!\!\right]}

\newcommand{\fun}[2]{\,{#1}\!\left[ {#2} \right]}

\newcommand{\real}[1]{\fun{\Re\mathfrak{e}}{#1}}

\newcommand{\unit}[1]{\,\mbox{#1}}

\newcommand{\Hz}{\unit{Hz}}
\newcommand{\kHz}{\unit{kHz}}

\newcommand{\meter}{\unit{m}}
\newcommand{\cm}{\unit{cm}}
\newcommand{\mm}{\unit{mm}}

\newcommand{\nm}{\unit{nm}}

\newcommand{\kg}{\unit{kg}}

\newcommand{\MW}{\unit{MW}}

\def\pint{\iint\limits_{\infty}}
\def\sint{\iint\limits_{\mbox{\tiny{surface}}}}
\def\vint{\iiint\limits_{\mbox{\tiny{optic}}}}

\usepackage{graphicx}

\begin{document}

\def\figWidth{3.3in}


\title{A General Approach to Optomechanical Parametric Instabilities}
\author{M.Evans$^1$,
 L.Barsotti$^1$,
 P.Fritschel$^1$}

\address{$^1$Massachusetts Institute of Technology, Cambridge, Massachusetts 02139, USA}


\begin{abstract}
We present a simple feedback description of parametric instabilities
 which can be applied to a variety of optical systems.
Parametric instabilities are of particular interest to
 the field of gravitational-wave interferometry where high mechanical
 quality factors and a large amount of stored optical power
 have the potential for instability.
In our use of Advanced LIGO as an example application,
 we find that parametric instabilities, if left unaddressed,
 present a potential threat to the stability of high-power operation.
\end{abstract}

\maketitle

\setcounter{MaxMatrixCols}{12}

\section{Introduction}

Though unseen in the currently operating
 first generation interferometric gravitational-wave antennae
 (e.g., GEO \cite{httpGEO}, TAMA \cite{httpTAMA},
  Virgo \cite{httpVirgo}, LIGO \cite{LIGO_RPP2009}),
 designers of second generation antennae
 may be faced with instabilities which result from the transfer
 of optical energy stored in the detector's Fabry-Perot cavities
 to the mechanical modes of the mirrors which form these cavity.

The field of gravitational wave interferometry was introduced to the
 concept of parametric instabilities (PI) couched in the language
 of quantum mechanics \cite{Braginsky2001331}.
We present a classical framework for PI which uses the audio-sideband
 formalism to represent the optical response of the interferometer \cite{Fritschel2001b,Regehr1995}.
This in turn allows the formalism to be extended to multiple interferometer configurations
 without the need to rederive the relevant relationships;
 an activity that has consumed considerable resources
 \cite{Gurkovsky200791,Gurkovsky2007177,Ju2006360,Strigin200710}.

\section{Parametric Instabilities}
\slabel{PI_intro}

The process which can lead to PI can be approached as a classical feedback effect
 in which mechanical modes of an optical system act on the
 electro-magnetic modes of the system via scattering,
 and the electro-magnetic modes act on the mechanical modes
 via radiation pressure (see figure \fref{LoopPI}).

\begin{figure}[h]
  \includegraphics[width=\figWidth]{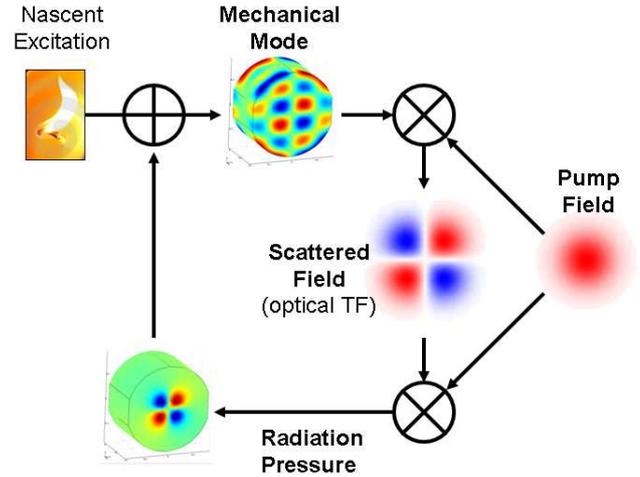}
  \caption{Parametric instabilities can be described as a classical feedback phenomenon.
  A given mechanical mode interacts with the pump field to scatter energy into higher order
  optical modes.
  This interaction is represented with $\otimes$ above.
  Its strength is given by the overlap integral of the mechanical mode,
   pump mode, and scattered field mode ($B_{m,n}$ in equation \eref{Bmn}).
  While circulating in the interferometer, the scattered field interacts with the
   pump field and mechanical mode via radiation pressure,
   introducing the overlap integral $\otimes$ a second time.}
  \flabel{LoopPI}
\end{figure}

We start by considering a single mechanical mode of one
 optic in the interferometer, and computing the parametric gain
 of that mode \cite{Kells2002326}.
The resonant frequency of this mode determines the frequency of
 interest for the feed-back calculation.\footnote{In general, all of the modes of all of the
 optics should be considered simultaneously, but with typical mode quality factors
 greater than $10^6$ it is very unlikely that two mechanical modes will participate
 significantly at the same frequency.}
The nascent excitation of this mechanical mode,
 possibly of thermal origin, begins the process by scattering
 light from the fundamental mode of the optical system into
 higher order modes (HOM).
The resulting scattered field amplitudes are
\begin{equation}
\elabel{E_scatter}
E_{scat,n} = \frac{2 \pi i}{\lambda_0} A_m E_{pump} B_{m, n}
\end{equation}
 where $\lambda_0$ is the wavelength and $E_{pump}$ the amplitude,
 of the ``pump field'' at the optic's surface,
 $A_m$ is the amplitude of the motion of the mechanical mode,
 and $B_{m, n}$ is its overlap coefficient with the $n^{th}$ optical mode.
The overlap coefficient results from an overlap integral of basis functions
 on the optic's surface
\begin{equation}
\elabel{Bmn}
B_{m, n} = \sint f_0 f_n (\vec{u}_m \cdot \hat{z}) \,d\vec{r}_{\perp}
\end{equation}
 where $\vec{u}_m$ is the displacement function of the mechanical mode,
 $f_0$ and $f_n$ are the field distribution functions for the pump field,
 typically a gaussian, and the $n^{th}$ HOM.
Each of these basis functions is normalized such that
\begin{eqnarray}
\vint |\vec{u}_m|^2 \,d\vec{r} & = & V \\
\pint |f_n|^2 \,d\vec{r}_{\perp} & = & 1
\end{eqnarray}
 where $V$ is the volume of the optic.\footnote{
A more general normalization of $\vec{u}_m$ which allows for non-uniform density
 would include the density function of the optic in the integral,
 with the result equal to the total mass.}

The interferometer's response to the scattered field can be computed in the
 modal basis used to express the HOMs via the audio sideband formalism.
The resulting transfer coefficients represent the gain and phase of the
 optical system to the scattered field as it travels from
 the optic's surface through the optical system and back.
The optical mode amplitudes of the field which returns to the optic's surface is
\begin{eqnarray}
\elabel{E_return}
E_{rtrn,n} & = & G_n E_{scat,n} \nonumber \\
 & = & \frac{2 \pi i}{\lambda_0} A_m E_{pump} G_n B_{m, n}
\end{eqnarray}
 where $G_n$ is the transfer coefficient from a field leaving
 the optic's surface to a field incident on and then reflected from the same surface.
The transfer coefficient $G_n$ is complex, representing the amplitude and phase of the
 optical system's response at the mechanical mode frequency.
Computation of $G_n$ for an optical system is discussed in
 more detail in section \sref{OptTC}.

The scattered field which returns to the optic closes the PI feedback loop
 by generating radiation pressure on that surface with a spatial profile
 that has some overlap with the mechanical mode of interest.
This radiation pressure force is given by
\begin{eqnarray}
F_{rad} & = & \frac{2}{c} E_{pump}^{\ast} \sum_{n = 0}^{\infty} B_{m, n} E_{rtrn, n} \nonumber \\
 & = & \frac{2 P}{c} \frac{2 \pi i}{\lambda_0} A_m \sum_{n = 0}^{\infty} G_n B_{m, n}^2
\elabel{F_rad}
\end{eqnarray}
 where $P = |E_{pump}|^2$, and $c$ is the speed of light.
The factor of 2 appears since the force is generated by the fields
 as they reflect from the surface.

Finally, the radiation pressure which couples into this mechanical
 mode adds to the source amplitude according to the transfer function
 of the mechanical system at its resonance frequency $\omega_m$,
\begin{eqnarray}
\elabel{A_prime}
\Delta A_{m} & = & \frac{-i Q_m}{M \omega_m^2} F_{rad} \nonumber \\
 & = & \frac{Q_m}{M \omega_m^2} \frac{4 \pi P}{c \lambda_0} A_m
  \sum_{n = 0}^{\infty} G_n B_{m, n}^2
\end{eqnarray}
 where $M$ is the mass of the optic
 and $Q_m$ the quality factor of the mechanical resonance.
The open-loop-gain of the PI feed back loop is therefore,
\begin{equation}
\elabel{PI_LoopGain}
\frac{\Delta A_{m}}{A_{m}}
 = \frac{4 \pi Q_m P}{M \omega_m^2 c \lambda_0} \sum_{n = 0}^{\infty} G_n B_{m, n}^2
\end{equation}

The parametric gain $\mathcal{R}$ is the real part of the open-loop-gain
\begin{eqnarray}
\elabel{R_m}
 \mathcal{R}_m & = & \real{\frac{\Delta A_{m}}{A_{m}}} \nonumber \\
 & = & \frac{4 \pi Q_m P}{M \omega_m^2 c \lambda_0}
  \sum_{n = 0}^{\infty} \real{G_n} B_{m, n}^2
\end{eqnarray}
 with the usual implication of instability if $\mathcal{R}_m > 1$,
 and optical damping if $\mathcal{R}_m < 0$.\footnote{
Appendix \sref{Comparison} relates this result to the results found
 in previous works.}

\section{Optical Transfer Coefficients}
\slabel{OptTC}

When the pump field is phase modulated by the mechanical mode of the optic,
 upper and lower scattering sidebands are produced.
In general, these scattering sidebands will have different optical transfer coefficients
 in the interferometer; $G_n^+$ and $G_n^-$.
The combination of scattering sidebands which leads to radiation pressure is
\begin{equation}
G_n = G_n^- - G_n^{+\ast}
\end{equation}
 (see appendix \sref{DeriveFrad} for derivation).
This section will describe a general method for computing $G_n^\pm$.

Given scattering matrices $\mathbb{S}_n^\pm$ which contain transfer coefficients
 for the $n^{th}$ HOM of the scattering sidebands from one point to the next in the
 optical system, we have
\begin{equation}
G_n^\pm = \vec{e}_x^{\hspace{2pt} T} \left( \mathbb{I} - \mathbb{S}_n^\pm \right)^{-1} \vec{e}_x
\end{equation}
 where the basis vector $\vec{e}_x$ is
 the $x^{th}$ column of the identity matrix $\mathbb{I}$,
 and $\vec{e}_x^{\hspace{2pt} T}$ is its transpose.
The index $x$ is used to select the field which reflects from optic of interest,
 as demonstrated in the following section.\footnote{
A general framework for constructing scattering matrices is described in \cite{Corbitt2005}
 and will not be reproduced here.}

\subsection{An Example: Fabry Perot Cavity}
\slabel{cavFP}

\begin{figure}[h]
  \includegraphics[width=\figWidth]{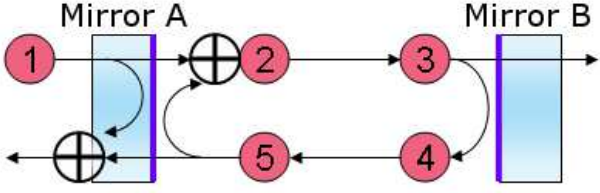}
  \caption{A simple Fabry Perot cavity.
  In the example caculation, fields are evaluated at each of the numbered circles.
  Field 4 is used to compute the optical gain of the scattered field produced by,
    and acting on, mirror B.}
  \flabel{OptFP}
\end{figure}

As a simple and concrete example, we apply the above formalism
 to a Fabry Perot cavity (FPC) of length $L$.
Figure \fref{OptFP} shows the configuration and indices for each of the fields in the FPC.

The scattering matrices for the upper and lower sidebands are
\begin{equation}
\mathbb{S}_n^\pm =
\begin{pmatrix}
0 & 0 & 0 & 0 & 0 \\
t_A & 0 & 0 & 0 & -r_A \\
0 & p_L^\pm & 0 & 0 & 0 \\
0 & 0 & -r_B & 0 & 0 \\
0 & 0 & 0 & p_L^\pm  & 0
\end{pmatrix}
\end{equation}
 where $t_A$ is the transmissivity for mirror A,
 $r_A$ and $r_B$ are the reflectivities of the mirrors,
 and $p_L^\pm = e^{i (\phi_n \pm \omega_m L / c)}$ is the propagation operator.
The reflectivity and transmissivity used in $\mathbb{S}_n^\pm$
 are amplitude values and are related to a
 mirror's power transmission by $t = \sqrt{T}$ and $r = \sqrt{1 - T}$.
The propagation phase depends on the phase of the $n^{th}$ HOM, $\phi_n$,
 and the scattering sideband frequency $\pm \omega_m$.

If we wish to evaluate $\mathcal{R}_m$ for a mode of mirror B, we would use
\begin{equation}
\vec{e}_4 =
\begin{pmatrix}
0 \\ 0 \\ 0 \\ 1 \\ 0
\end{pmatrix}
\end{equation}
 to select its reflected field, number 4 in figure \fref{OptFP}.
To arrive at a numerical result for $\mathcal{R}_m$,
 we adopt the following parameters
\begin{align*}
P &= 1 \MW  & \lambda_0 &= 1064 \nm \\
T_A &= 0.014 & T_B &= 10^{-5}\\
L &= 3994.5 \meter & M &= 40 \kg
\end{align*}
 which are representative of an Advanced LIGO arm cavity
 operating at full power.
For the moment, we will consider a single optical mode,
 the Hermite-Gauss TEM11 mode, and a single mechanical mode
\begin{align*}
Q_m &= 10^7 & \omega_m &= 2 \pi \times 29950 \Hz \\
B_{m,HG11} &= 0.21 & \phi_{HG11} &= -5.434
\end{align*}
 both of which are shown in figure \fref{FPmodes}.

\begin{figure}[h]
  \includegraphics[width=\figWidth]{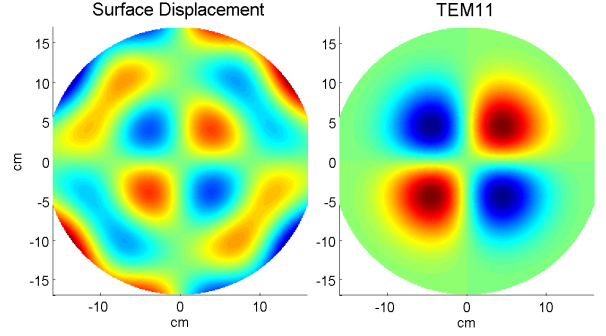}
  \caption{A mechanical mode of an Advanced LIGO test mass near 30kHz
   and a HG TEM11 optical mode.
   For the mechanical mode, surface displacement amplitude normal to the surface,
   $\vec{u}_m \cdot \hat{z}$, is shown.
   For the optical mode, the basis function $f_{HG11}$ amplitude is shown.
   In both cases, red is positive, blue is negative and green is zero.
   The X and Y-axes on both plots are in centimeters.}
  \flabel{FPmodes}
\end{figure}

For this set of values, we evaluate the parametric gain
\begin{align*}
G_{HG11}^+ &= 0.554 + i 2.72 & G_{HG11}^- &= 0.502 + i 0.03 \\
\Rightarrow G_{HG11} &= -0.052 + i 2.75 & \mathcal{R}_{m,HG11} &= -6.5 \times 10^{-4}
\end{align*}
 to find it is small and negative.

Allowing the mechanical mode frequency to artificially vary from $20\kHz$ to $50\kHz$
 we can plot $\mathcal{R}_m$ as a function of $\omega_m$.
The result is shown in figure \fref{FPgains11}.
The resonance at $27.4\kHz$ is the upper scattering sideband of the TEM11 mode
 which has negative parametric gain, indicating optical damping.
At $47.7\kHz$ the lower scattering sideband of the TEM11 mode resonates,
 this time resulting in positive feedback,
 but not enough to produce instability.

\begin{figure}[h]
  \includegraphics[width=\figWidth]{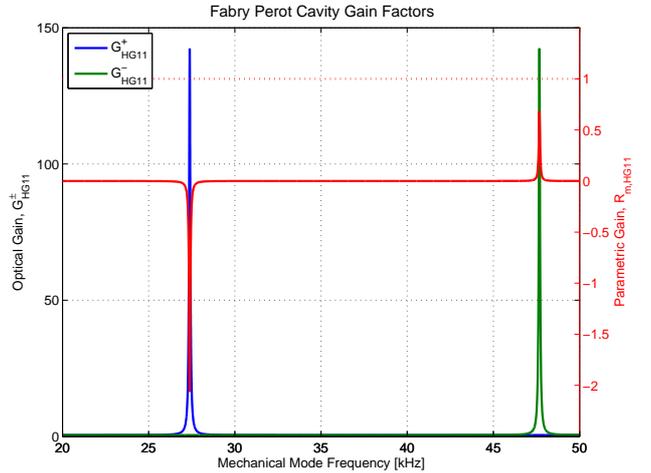}
  \caption{Optical gains $G_{HG11}^+$ and $G_{HG11}^-$, and parametric gain $\mathcal{R}_{m,HG11}$
   are shown as a function of mechanical mode frequency.
The resonance of the upper scattering sideband at $27.4\kHz$ has negative parametric gain,
 indicating optical damping.
The lower scattering sideband resonance at $47.7\kHz$ has positive gain,
 but does not produce instability.}
  \flabel{FPgains11}
\end{figure}

Thus far only one optical mode and only one mechanical mode have been considered.
Extending the computation to higher order optical modes requires that we elaborate
 our expression for $\phi_n$ to include the Gouy phase.
For an arbitrary Hermite-Gauss mode
\begin{equation}
\phi_n = \phi_0 - O_n \phi_G
\end{equation}
 where $\phi_0$ is the propagation phase of the TEM00 mode,
 and $O_n$ is the mode order of the $n^{th}$ HOM.

Considering other mechanical modes is a matter of computing the mode shapes and frequencies
 for the mirrors which make up the FPC;
 we use those of an Advanced LIGO test-mass.\footnote{
The Advanced LIGO test-mass mechanical modes used in this and the following examples
 are the result of finite element modeling.
For a discussion of numerical and analytic methods for calculating test-mass mechanical
 modes, see \cite{Strigin20086305}.}
The result of the full calculation of $\mathcal{R}_m$ for all mechanical modes
 between $10\kHz$ and $90\kHz$, including HOMs up to $9^{th}$ order,
 is shown in figure \fref{FPparaGains}.

\begin{figure}[h]
  \includegraphics[width=\figWidth]{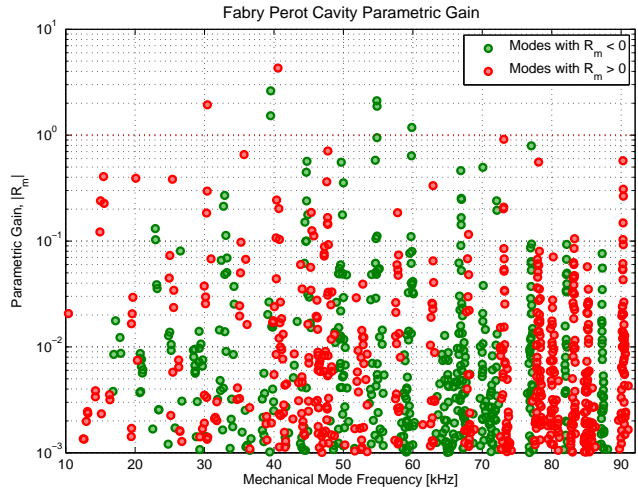}
  \caption{Parametric gains $\mathcal{R}_m$
   are shown for mechanical modes between $10\kHz$ and $90\kHz$.
   Red circles mark modes with positive parametric gain,
    while green circles mark those with negative gain.
   This calculation uses HOMs up to $9^{th}$ order,
    but does not include clipping losses,
    discussed in section \sref{Clipping}.}
  \flabel{FPparaGains}
\end{figure}

\section{Clipping Losses}
\slabel{Clipping}

Thus far we have ignored losses in the optical system.
A lower limit on the losses is given by the loss of power
 due to the finite size of the optics, known as clipping loss.
For low-order optical modes, these losses are usually insignificant
 by design, but losses can strongly impact the parametric gain
 when the contribution from high-order optical modes is dominant.

In optical systems such as gravitational-wave interferometers,
 in which the beam size on the optics is made as large as possible
 without introducing significant loss in the TEM00 mode,
 high-order modes tend to fall off the cavity optics.
Specifically, for an interferometer designed to have a few
 parts-per-million clipping losses for the TEM00 mode,
 contributions to $\mathcal{R}_m$ from modes of order $O_n \gtrsim 4$ are limited,
 and modes with $O_n \gtrsim 9$ are insignificant.

A more complete description of losses due to apertures includes
 diffraction effects, but this requires a more complex
 and interferometer dependent calculation.
Even better is to use the eigenmodes of the full interferometer,
 and their associated losses, rather than the Hermite-Gauss basis.
This level of detail may not be rewarded,
 however, since modes which differ significantly from their
 Hermite-Gauss partners do so as a result of significant losses,
 which in turn make them irrelevant to PI.

\subsection{An Example: Advanced LIGO}
\slabel{aLIGO}

\begin{figure}[h]
  \includegraphics[width=\figWidth]{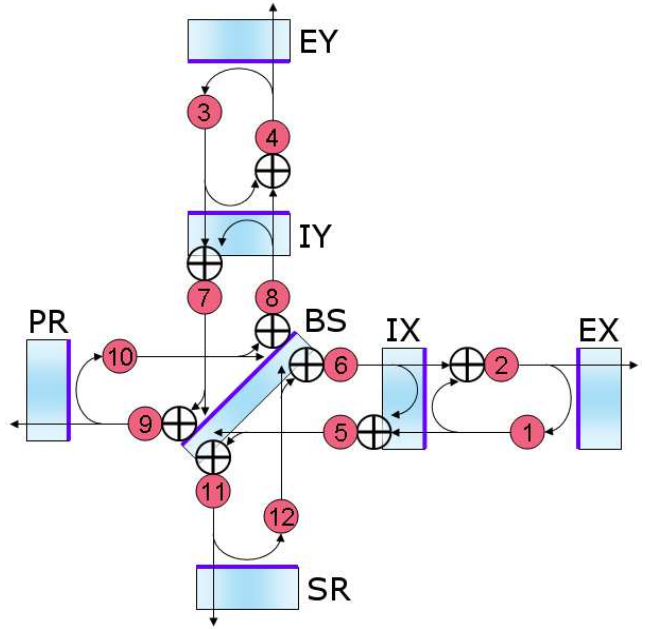}
  \caption{Fields in a power and signal recycled Fabry Perot Michelson.
  This optical configuration is common to many of the $2^{nd}$ generation
  gravitational-wave detectors.}
  \flabel{OptSR}
\end{figure}

As a more interesting example, we apply the above formalism to an
 Advanced LIGO interferometer.
Figure \fref{OptSR} shows the layout of the optical system and the assignment
 of field evaluation points (FEPs).
In this case care has been taken to minimize the number of FEPs and to follow
 each one with a propagation operation.
In this way we can number the propagation distances $L_x$ according to
 their associated FEP, and the propagation operations become
 $p_{n,x}^\pm = e^{i (\phi_{n,x} \pm \omega_m L_x / c)}$.

The scattering matrices for this interferometer can be split into
 a diagonal propagation matrix populated by $p_{n,x}^\pm$,
 and a mirror matrix populated by reflectivity and transmissivity
 coefficients (essentially one $r$ and one $t$ in each column),
 as follows
\begin{equation}
\elabel{S_n}
\def\mz{\cdot}
\mathbb{S}_n^\pm = \mathbb{M}_n \, \mathbb{P}_n^\pm
 \hspace{8ex}
\mathbb{P}_n^\pm =
\left(
\begin{array}{*{3}c}
p_{n,1}^\pm & \mz & \cdots \\
\mz & p_{n,2}^\pm & \\
\vdots & & \ddots
\end{array}
 \right)
\end{equation}
\begin{equation*}
\def\mz{\cdot}
\def\mm{{\scriptscriptstyle -}}
\renewcommand{\arraycolsep}{-1pt}
\renewcommand{\arraystretch}{0.7}
\mathbb{M} \hspace{-1pt} = \hspace{-2pt}
\left(
\begin{array}{*{12}c}
 \mz & \mm r_{EX} & \mz & \mz & \mz & \mz & \mz & \mz & \mz & \mz & \mz & \mz \\
\mm r_{IX} & \mz & \mz & \mz & \mz & t_{IX} & \mz & \mz & \mz & \mz & \mz & \mz \\
 \mz & \mz & \mz & \mm r_{EY} & \mz & \mz & \mz & \mz & \mz & \mz & \mz & \mz \\
 \mz & \mz & \mm r_{IY} & \mz & \mz & \mz & \mz & t_{IY} & \mz & \mz & \mz & \mz \\
t_{IX} & \mz & \mz & \mz & \mz & r_{IX} & \mz & \mz & \mz & \mz & \mz & \mz \\
 \mz & \mz & \mz & \mz & \mz & \mz & \mz & \mz & \mz & t_{BS} & \mz & r_{BS} \\
 \mz & \mz & t_{IY} & \mz & \mz & \mz & \mz & r_{IY} & \mz & \mz & \mz & \mz \\
 \mz & \mz & \mz & \mz & \mz & \mz & \mz & \mz & \mz & \mm r_{BS} & \mz & t_{BS} \\
 \mz & \mz & \mz & \mz & t_{BS} & \mz & \mm r_{BS} & \mz & \mz & \mz & \mz & \mz \\
 \mz & \mz & \mz & \mz & \mz & \mz & \mz & \mz & \mm r_{PR} & \mz & \mz & \mz \\
 \mz & \mz & \mz & \mz & r_{BS} & \mz & t_{BS} & \mz & \mz & \mz & \mz & \mz \\
 \mz & \mz & \mz & \mz & \mz & \mz & \mz & \mz & \mz & \mz & \mm r_{SR} & \mz
\end{array}
 \right)
\end{equation*}

Losses can be included in the scattering matrix $\mathbb{S}_n^\pm$
 most generally by allowing $\mathbb{M}_n$ to vary for each HOM.
A simpler approach which is sufficient for clipping losses is
 to add a diagonal matrix $\mathbb{C}_n$ to equation \eref{S_n}
 which effectively adds loss to each propagation step
\begin{gather}
\def\mz{\cdot}
\mathbb{S}_n^\pm = \mathbb{M}  \, \mathbb{C}_n \, \mathbb{P}_n^\pm
 \hspace{8ex}
\mathbb{C}_n =
\left(
\begin{array}{*{3}c}
t_{n,1} & \mz & \cdots \\
\mz & t_{n,2} & \\
\vdots & & \ddots
\end{array}
 \right)
\end{gather}
  where
\begin{equation}
t_{n,x} = \sqrt{\, \sint f_n^2 \, d\vec{r}_{\perp}}
\end{equation}
 is the amplitude transmission of the aperture associated with each propagation step.
For this example, a $17\cm$ aperture is assumed for all of the optics,
 except the beam-splitter for which we assume a $13.3\cm$ aperture.\footnote{
The beam-splitter aperture includes the aperture presented by the
 electro-static actuators on IX and IY.}

In order to compute the parametric gain as a function of mechanical mode frequency,
 as in figure \fref{FPgains11} for the FPC example, we use the same values as above
 \begin{gather*}
P = 1 \MW  \quad  \lambda_0 = 1064 \nm \\
M = 40 \kg \quad Q_m = 10^7
\end{gather*}
 and add the transmission of the new optics
\begin{gather*}
T_{IX} = T_{IY} = 0.014 \quad  T_{EX} = T_{EY} = 10^{-5}\\
T_{PR} = 0.03 \quad  T_{SR} = 0.2 \quad  T_{BS} = 0.5
\end{gather*}
 new lengths
\begin{gather*}
L_{\{1,2,3,4\}} = 3994.5 \meter \\
L_{\{5,6\}} = 4.85 \meter \quad L_{\{7,8\}} = 4.9 \meter \\
L_{\{9,10\}} = 52.3 \meter \quad L_{\{11,12\}} = 50.6 \meter
\end{gather*}
 and phases
\begin{gather*}
\phi_{0,\{1-8,11,12\}} = 0 \quad \phi_{0,\{9,10\}} = \pi / 2 \\
\phi_{G,\{1,2,3,4\}} = 2.72 \quad \phi_{G,\{5,6,7,8\}} = 0 \\
\phi_{G,\{9,10\}} = 0.44 \quad \phi_{G,\{11,12\}} = 0.35
\end{gather*}
 which represent $156^\circ$ of Gouy phase in the arms,
 $25^\circ$ in the power recycling cavity and
 $20^\circ$ in the signal recycling cavity.
The results are similar to the FPC alone, with negative gain near $27.4\kHz$
 and positive gain near $47.7\kHz$.
The addition of the rest of the interferometer, however,
 leads to a narrow regions of high parametric gain for the HG11 mode, see figure \fref{ALIGOgains11}.

\begin{figure}[ht]
  \includegraphics[width=\figWidth]{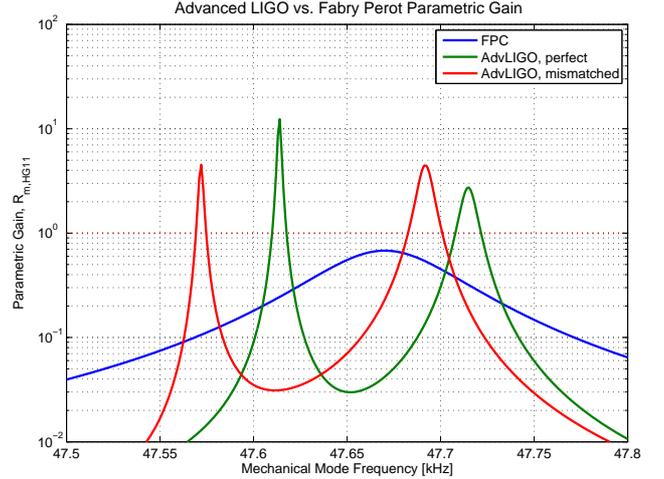}
  \caption{Comparison of $\mathcal{R}_{m, HG11}$ for a Fabry Perot cavity
   and Advanced LIGO are shown as a function of mechanical mode frequency.
  The Advanced LIGO computation is shown twice,
   the green curve includes clipping losses, but has perfectly matched arm cavities,
   the red curve adds a $0.1\%$ mismatch between the arm cavity Gouy phases.
  The calculation is limited to the modes shown in figure \fref{FPmodes},
   with the mechanical mode frequency artificially
   adjusted to highlight the resonance near $47.7\kHz$.}
  \flabel{ALIGOgains11}
\end{figure}

\begin{figure}[ht]
  \includegraphics[width=\figWidth]{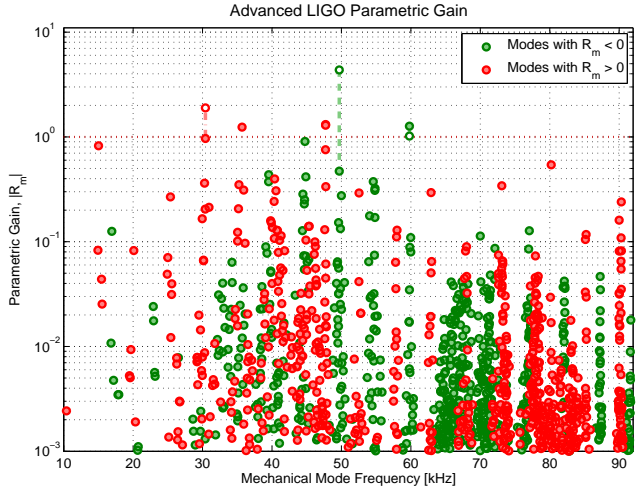}
  \caption{Parametric gain for all modes of an Advanced LIGO
   test-mass between $10\kHz$ and $90\kHz$.
   To show the effect of clipping losses, a few of the modes
   which have significant gain in the absence of clipping are also included.
Empty circles represent parametric gains computed without clipping.
They are attached to their clipped partners, represented by filled circles, with dashed lines.}
  \flabel{ALIGOparaGains}
\end{figure}

As noted in \cite{Strigin200710}, realistically imperfect matching of interferometer optics
 can significantly change the parametric gains of the system.
Our formalism can be made to reproduce this result by allowing the Gouy phase
 in one arm of the interferometer to differ from that of the other arm,
\begin{equation*}
\phi_{n,\{3,4\}} = (1 + \epsilon) \phi_{n,\{1,2\}}
\end{equation*}
 where $\epsilon_Y$ is the fractional departure of $\phi_{n,\{3,4\}}$ from their nominal value.
For our Advanced LIGO example,
 where radius of curvature errors of a few meters are expected,
 we take $\epsilon = 10^{-3}$.
This tiny difference is sufficient to move the sharp features associated with a perfectly
 matched interferometer by more than their width, thereby changing the result of any PI
 calculation (see figure \fref{ALIGOgains11}).  These results are similar to those in
 \cite{Gras2009CQG}, though their model of the power-recycling cavity was somewhat less general.

Finally, the full calculation for Advanced LIGO is plotted in figure \fref{ALIGOparaGains}.
To show the effect of clipping losses,
 modes which have $\mathcal{R}_n > 1$ in the absence of clipping
 are shown and connected to their clipped partners.

\section{Worst Case Analysis}
\slabel{WorseCase}

Evaluating the impact of PI on a gravitational-wave interferometer
 is complicated by the sensitivity of the result to small changes
 in the model parameters.
In particular, uncertainty in the radii of optics used in the
 arm cavities lead to changes the Gouy phases which,
 while quite small, are sufficient to move the optical
 resonances of the cavity by more than their width.
Similarly, mechanical mode frequencies produced analytically or by
 finite element modeling may not match the real articles due
 to small variations in materials, assembly, and ambient temperature.
This section describes a robust means of estimating the
 ``worst case scenario'' for a given interferometer.

A simple approach to the ``worst case'' problem is to compute the parametric
 gain of each mode for multiple sets of plausible interferometer
 parameters.
Varying all of the parameters is impractical and unnecessary
 as the results are primarily sensitive to the relative frequencies
 of the mechanical modes and the optical resonances in high-finesse
 cavities.

In the case of Advanced LIGO, explored in section \sref{aLIGO} above,
 it is sufficient to vary the Gouy phases in the arm cavites by
 a $5 \, \times \, 10^{-3}$, and the phases in the recycling cavities
 by a couple of degrees.
We proceed with a Monte-Carlo type analysis in which we randomly
 vary the Gouy phases around their nominal values.
We repeat the process for 120 thousand trials,
 then set an upper-limit on $\mathcal{R}_n$ for each mode at the
 lowest value greater than $99\%$ of the results
 (see figure \fref{ALIGOworstCase}).\footnote{
 To speed the computation, only optical modes
 with some overlap, $B_{m,n}^2 > 10^{-3}$,
 and modest clipping losses, $t_{n,1}^2 > 0.7$, are considered.}
This provides us with a trial number insensitive statistic,
 the accuracy of which is limited primarily by the
 fidelity of our model.

We find that Advanced LIGO faces the possibility of a few unstable modes.
Among the 120 thousand cases considered, the mean value of the maximum
 $\mathcal{R}_n$ among all modes was 5.8 and $99\%$ of cases had
 a maximum parametric gain value less than 45.
Considering each mechanical mode independently, there are 32
 modes which have the potential to be unstable and that all of the
 highly unstable modes are between $15\kHz$ and $50\kHz$.
Taking into account that 4 test-masses make up the Advanced LIGO detector,
 we find the mean number of unstable modes to be 10 (2.5 per test-mass),
 with $99\%$ of cases having 6 or fewer unstable modes per test-mass.

One must keep in mind, however, that many of the parameters used in
 this model are adjustable (e.g., power level in the interferometer,
 mirror temperature and thus mechanical mode frequency and radius
 of curvature, etc.) and others are speculative
 (e.g., the quality factor of a mechanical mode may depend
 strongly on its suspension \cite{Logan92, Rowan1998}).
Since the parametric gain scales directly with both power in the interferometer
 and with mechanical mode Q, we have chosen round values for these parameters
 which can be refined as higher fidelity numbers become known.

\begin{figure}[ht]
  \includegraphics[width=\figWidth]{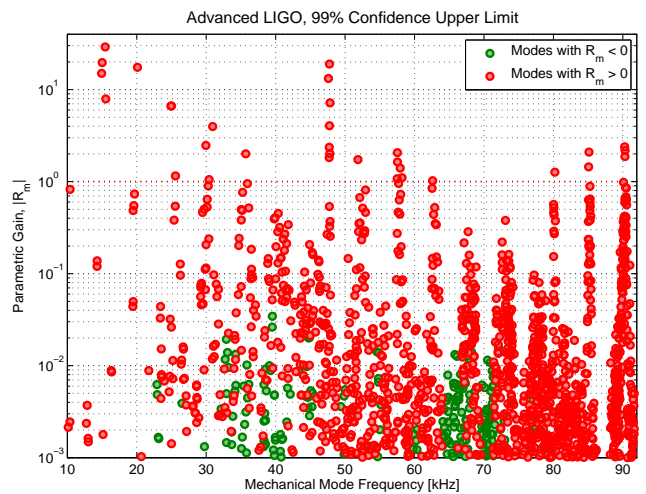}
  \caption{Worst case parametric gain for all modes of an Advanced LIGO
   test-mass between $10\kHz$ and $90\kHz$.
   There are 32 potentially unstable modes,
   and more than 200 modes with $\mathcal{R}_n > 0.1$.}
  \flabel{ALIGOworstCase}
\end{figure}

\section{Conclusions}
\slabel{Conclusions}

Parametric instabilities are of particular interest to
 the field of gravitational-wave interferometry where high mechanical
 quality factors and a large amount of stored optical power
 has the potential for instability.
We depart from previous work by constructing a flexible
 analysis framework which can be applied to a variety of optical systems.
Though our examples use a Hermite-Gaussian modal basis to describe the
 optical fields, this formalism can be implemented using the
 modal basis best suited to the optical system at hand.

In our use of Advanced LIGO as an example application,
 we find that parametric instabilities, if left unaddressed,
 present a potential threat to the stability of high-power operation.
We hope that future work on solutions to parametric instabilities will be
 guided by these results.

\newpage
\appendix

\section{Comparison with Previous Works}
\slabel{Comparison}


The notation used herein has been chosen to match
 the seminal work by Braginsky where possible.
To relate this work to \cite{Braginsky2001331}, which considered a single Fabry-Perot cavity,
 we need the following equalities
\begin{align*}
\frac{2 P}{c} & = \frac{\mathcal{E}_0}{L} & B_{m, n}^2 & = \Lambda \\
\frac{2 \pi}{\lambda_0} \Re [G_n] & = \frac{2 Q_1}{L (1 + \Delta \omega^2 / \delta_1^2)} &
 M & = m
\end{align*}
 where symbols on the left side of each equation are defined herein,
 and the symbols on the right are used in their equations 3, 4, 8 and 9.
Braginsky uses $m$ for the total mass of the optic, but here that symbol is
  used as the mechanical mode index, so we use $M$ instead.
For reference, their equation 4 is
\begin{equation*}
\mathcal{R}_0 = \frac{2 \mathcal{E}_0 Q_1 Q_m}{m L^2 \omega_m^2}
\end{equation*}
 such that our $\mathcal{R}$ is the same as the left-hand side of their equation 8.
That is,
\begin{equation*}
 \mathcal{R}_m = \frac{\mathcal{R}_0 \Lambda}{1 + \Delta \omega^2 / \delta_1^2}
\end{equation*}
 where the symbols on the right are those of Braginsky.
One should note that Braginsky only considered one mechanical mode
 and one optical mode, so his subscript 1 replaces our $m,n$
 and the summation over modes is absent.
Similarly, $B_{m, n}$ in this work is the same as $B$ in \cite{Kells2002326}.

\section{Detailed Derivation of $F_{rad}$}
\slabel{DeriveFrad}

We start by examining more carefully equation \eref{E_scatter}
 in which the scattered field is generated by phase modulation
 of the pump field.
A complete description of the pump field reflected from the optic is
\begin{equation}
\elabel{Field_pump}
\mathcal{E}_{pump}(r_\perp) = E_{pump} f_0(r_\perp) \, e^{i \omega_0 t}
\end{equation}
 where $f_0(r_\perp)$ is the normalized basis function of the pump field,
 and $\omega_0 = 2 \pi c / \lambda_0$ is the frequency of the pump field.
Mechanical motion of the reflecting surface, with a spatial profile
 and frequency defined by a resonant mode of the optic
\begin{equation}
\elabel{A_mech}
\mathcal{Z}_m(r_\perp) = A_m (\vec{u}_m(r_\perp) \cdot \hat{z}) \, e^{i \omega_m t}
\end{equation}
 results in phase modulation of the pump field.
The scattered field is expressed to first order as a sum of HOM fields
\begin{equation}
\elabel{Field_scatter}
\mathcal{E}_{scat}(r_\perp) = \sum_{n = 0}^{\infty} E_{scat,n} f_n(r_\perp)
 \, e^{i \omega_0 t} (e^{i \omega_m t} + e^{-i \omega_m t})
\end{equation}
 with HOM basis functions $f_n(r_\perp)$, and HOM amplitudes
\begin{equation}
\elabel{E_scat_n}
E_{scat,n} = \frac{2 \pi i}{\lambda_0} A_m E_{pump} B_{m,n}.
\end{equation}
Note that in equation \eref{Field_scatter} there are 2 frequency components produced;
 these are the upper and lower scattering sidebands.

The radiation pressure which couples the scattered field into mechanical motion
 is given by
\begin{equation}
\mathcal{P}_{rad}(r_\perp) = \frac{2}{c} \,
 \mathcal{E}_{refl}(r_\perp) \mathcal{E}_{refl}^\ast(r_\perp)
\end{equation}
 where the reflected field is
\begin{equation}
\elabel{Field_reflected}
\mathcal{E}_{refl}(r_\perp) = \mathcal{E}_{pump}(r_\perp)
 + \mathcal{E}_{rtrn}^+(r_\perp) + \mathcal{E}_{rtrn}^-(r_\perp)
\end{equation}
 and $\mathcal{E}_{rtrn}^\pm$ are the return fields,
 with HOM amplitudes given by applying $G_n^\pm$ to the scattered field.
By defining $\mathcal{E}_{rtrn}^\pm$ as part of the reflected field,
 $G_n^\pm$ represents the closed-loop gain of each scattering field.

The component of radiation pressure which is at the mechanical mode frequency is
\begin{align}
\mathcal{P}_{rad}(r_\perp,\omega_m) &
= \int_{0}^{t \gg 1 / \omega_m} \hspace{-6ex} e^{i \omega_m t} \, \mathcal{P}_{rad}(r_\perp) \, dt\\
= \frac{2}{c} & \left( E_{pump}^\ast f_0(r_\perp)
 \sum_{n = 0}^{\infty} E_{rtrn,n}^- f_n(r_\perp) \right. \nonumber \\
 & \left. + E_{pump} f_0(r_\perp)
  \sum_{n = 0}^{\infty} E_{rtrn,n}^{+\ast} f_n(r_\perp) \right) \nonumber \\
= \frac{2}{c} & \left( E_{pump}^\ast f_0(r_\perp)
 \sum_{n = 0}^{\infty} (G_n^- E_{scat,n}) f_n(r_\perp) \right. \nonumber \\
 & \left. + E_{pump} f_0(r_\perp)
  \sum_{n = 0}^{\infty} (G_n^+ E_{scat,n})^{\ast} f_n(r_\perp) \right). \nonumber
\end{align}
Integrating over the surface to get the force which couples
 into the mechanical mode of interest
 simplifies the notation somewhat to
\begin{align}
F_{rad} = \frac{2}{c} & \left( E_{pump}^\ast
 \sum_{n = 0}^{\infty} (G_n^- E_{scat,n}) B_{m,n} \right. \nonumber \\
 & \left. + E_{pump}
  \sum_{n = 0}^{\infty} (G_n^+ E_{scat,n})^{\ast} B_{m,n} \right) \\
= \frac{2 P}{c} & \frac{2 \pi i}{\lambda_0} A_m
 \sum_{n = 0}^{\infty} \left( G_n^- - G_n^{+\ast} \right) B_{m,n}^2
\end{align}
 where the last step comes from substitution using equation \eref{E_scat_n}.
We recover equation \eref{F_rad} by setting
\begin{equation}
G_n = G_n^- - G_n^{+\ast}.
\end{equation}

\bibliography{ClassicalPI}

\end{document}